# TITLE

Analyses of 'change scores' do not estimate causal effects in observational data

# AUTHORS

*Peter WG Tennant[1,2,3], Kellyn F Arnold[1,2], George TH Ellison[1,2], Mark S Gilthorpe[1,2,3]

[1]Leeds Institute for Data Analytics, University of Leeds, Leeds, UK; [2]School of Medicine, University of Leeds, Leeds, UK; [3]The Alan Turing Institute, London, UK

*Corresponding Author: Peter WG Tennant. Leeds Institute for Data Analytics, University of Leeds, Level 11 Worsley Building, Clarendon Way, Leeds, LS2 9NL, UK. Email: P.W.G.Tennant@leeds.ac.uk. Tel: +44 (0)113 278 3226.

# ATTRIBUTES

*Word count (Manuscript, excluding references):* 3522

*Word count (Abstract):* 206

*References:* 16 (+ 5 supplementary)

*Tables:* 1 (+ 1 supplementary)

*Figures:* 3

# ABSTRACT

**Background**

In longitudinal data, it is common to create 'change scores' by subtracting measurements taken at baseline from those taken at follow-up, and then to analyse the resulting 'change' as the outcome variable. In observational data, this approach can produce misleading causal effect estimates. The present article uses directed acyclic graphs (DAGs) and simple simulations to provide an accessible explanation of why change scores do not estimate causal effects in observational data.

**Methods**

Data were simulated to match three general scenarios where the variable representing measurements of the outcome at baseline was a 1) competing exposure, 2) confounder, or 3) mediator for the total causal effect of the exposure on the variable representing measurements of the outcome at follow-up. Regression coefficients were compared between change-score analyses and DAG-informed analyses.

**Results**

Change-score analyses do not provide meaningful causal effect estimates unless the variable representing measurements of the outcome at baseline is a competing exposure, as in a randomised experiment. Where such variables (i.e. baseline measurements of the outcome) are confounders or mediators, the conclusions drawn from analyses of change scores diverge (potentially substantially) from those of DAG-informed analyses.

**Conclusions**

Future observational studies that seek causal effect estimates should avoid analysing change scores and adopt alternative analytical strategies.

# KEY MESSAGES

- 'Change scores' provide a simple summary measure of the average change in a variable between two time points; they are commonly used when analysing 'change' in an outcome with respect to a baseline exposure.

- Analyses of outcome change scores do not estimate causal effects except under randomised experimental conditions; in some (non-randomised) situations, the implied 'effect' may be of opposite sign to the total causal effect.

- Future observational studies that seek causal effect estimates should avoid analysing outcome change scores and adopt alternative analytical strategies; studies that have conducted analyses of outcome change scores should be viewed with caution and their recommendations revisited.

# INTRODUCTION

Studies of change are a cornerstone of research in the health sciences. Understanding the natural history of disease, and in turn predicting prognoses, are of enormous interest to physicians and patients alike. Analyses of 'change' are, however, deceptively complex in observational data. One of the most common, yet poorly recognised, challenges stems from the use and interpretation of 'change scores'.

Change scores (e.g. $\Delta Y = Y_1 - Y_0$), also known as 'difference scores', 'gain scores', and 'change-from-baseline variables', are composite variables that have been constructed from repeated measures of a single parent variable ($Y$) by subtracting a subsequent measure of the parent ($Y_1$, 'follow-up') from a prior measure ($Y_0$, 'baseline'). The resulting composite variable retains information from both of its determining parents and hence will share a tautological association with either if analysed by regression or correlation.[1] This was first recognised by Oldham in 1962, who demonstrated that an association averaging $r = \pm 1/\sqrt{2}$ occurs between either of the parent variables (i.e. $Y_0$ or $Y_1$) and their difference (i.e. $Y_1 - Y_0$) if both have similar variances but are otherwise unrelated.[2] This phenomenon explains the 'law of initial value' as a consequence of the sign disagreement between the baseline parent variable ($Y_0$) and its transformation in the composite change-score ($-Y_0$), and is distinct from regression-to-the-mean.[1]

Relatively few analyses of change scores, however, involve straightforward tautological associations. More often, change scores are used as outcome variables in relation to a separate baseline treatment or exposure $X_0$ (e.g. 'How do beta-blockers affect change in blood pressure?'). One of the most widely recognised issues in this context is the discordance between change-score analyses (i.e. where the outcome change-score $\Delta Y$ is regressed on the baseline exposure $X_0$) and analyses of covariance (ANCOVA; i.e. where the follow-up outcome $Y_1$ is regressed on the baseline exposure $X_0$ and 'adjusted for' the baseline outcome $Y_0$).[3,4] For example, Senn (2006) and Van Breukelen (2006) found that change-score analyses and ANCOVA provide similar and unbiased estimates when the exposure is randomised but provide 'contradictory results' when the exposure is not randomised.

Frederick Lord's eponymous paradox centres on this same 'contradiction' and the lack of an obvious 'correct' answer.[5]

Although studies of change are extremely common, the concept of change – and the use of change scores as a putative measure thereof – has received relatively limited formal consideration within a causal framework. Causal diagrams such as directed acyclic graphs (DAGs) provide a useful framework for understanding some challenges associated with observational data analysis, but they have not often been used to consider analyses of change scores specifically. Of the exceptions, Glymour et al. (2006) focussed on the role of measurement error, arguing that analyses of outcome change scores provide unbiased causal effect estimates in some cases, but that error can be introduced by conditioning on the baseline outcome.[6] Conversely, Shahar and Shahar (2010) argue that change scores are 'not of causal interest' and that 'modelling the change between two time points is justified only in a few situations'.[7]

The present article aims to provide an accessible explanation of why change scores cannot be used to estimate causal effects in observational (i.e. non-randomised) data.

## CHANGE SCORES DO NOT REPRESENT EXOGENOUS CHANGE

In this section, we consider the concept of 'change' using DAGs. We focus on 'exogenous change' in an outcome variable ($Y$), which represents the structural (i.e. non-random) component of the follow-up outcome ($Y_1$) that has not been determined at baseline ($Y_0$) and can therefore potentially still be modified after baseline.

DAGs are semi-parametric graphical representations of hypothesised causal relationships between variables.[8] Variables (depicted as nodes) are connected by unidirectional arcs (depicted as arrows), which indicate the presence and direction – though neither the nature nor magnitude – of each hypothesised causal relationship. A path is a collection of one or more arcs that connects two nodes, and a causal path is one where all constituent arcs flow in the same direction. No variable can cause itself. By convention, we depict deterministic variables as double-outlined nodes.[9]

We first consider the simple example of repeated measures of an outcome variable ($Y$) which only fluctuate due to randomness ($R$) (Figure 1A). Values of the follow-up ($Y_1$) are entirely determined by the baseline ($Y_0$) plus the random features at follow-up ($R_1$). In this scenario, $Y_1$ cannot be modified except by modifying $Y_0$; no exogenous change exists. This is obvious in repeated measures of a fixed variable, such as height in healthy middle-aged adults. Although each individual's height values $Y_0$ and $Y_1$ would likely differ slightly due to the random features at baseline ($R_0$) and follow-up ($R_1$), this only dilutes the observed relationship between $Y_0$ and $Y_1$. In the population, there would be no *overall* change in the average values of height at baseline and follow-up, and this would be correctly reflected by a change score with a mean of zero (Figure 1A+).

*[Insert Figure 1 here]*

The same causal scenario (i.e. Figure 1A) could also describe repeated measures of a *dynamic* variable, whereby follow-up values are *fully determined* by baseline values via an algebraic function. As an example, consider the total expected number of radioactive particles $Y$ in a sample of (non-depleted) uranium rods at some future point in time ($Y_1$), which may be estimated without bias from the current observed number of radioactive particles ($Y_0$) by the Universal Law of Radioactive Decay.[10] The total observed value of $Y$ would irrefutably change between $Y_0$ and $Y_1$, and each individual uranium rod would have a negative change score (the magnitude of which would increase with the size of $Y_0$). Nevertheless, no exogenous change exists; as previously, $Y_1$ cannot be modified except by modifying $Y_0$.

Finally, we consider a more realistic dynamic variable ($Y$), whose future values ($Y_1$) are only partly determined by the past values ($Y_0$), with the remainder determined by random features ($R_1$) plus other exogenous change ($C_1$) (Figure 1B). Here, $C_1$ represents all *non-random* change in $Y$ that is not pre-determined by $Y_0$, and so the concept of exogenous change can thus be considered an average of all the processes in $C_1 \rightarrow Y_1$. In reality, $C_1$ is an unmeasurable, latent variable whose value is only defined once the point of follow-up is fixed. Thus, the exogenous change between two time points is fundamentally encapsulated within $Y_1$.

# ISOLATING EXOGENOUS CHANGE WITH RESPECT TO A BASELINE EXPOSURE

The causal effect of a baseline exposure $X_0$ on 'change' in $Y$ hence corresponds to the effect of $X_0$ on 'exogenous change' in $Y$, i.e. the structural part of $Y_1$ that has not already been determined by $Y_0$. This quantity can be expressed as the effect of $X_0$ on $Y_1|Y_0$, and may be estimated by constructing, for example, a regression model of the form $\widehat{Y_1} = \hat{\alpha}_0 + \hat{\alpha}_1 X_0 + \hat{\alpha}_2 Y_0$. We refer to this analysis as the **follow-up adjusted for baseline analysis**, where $\hat{\alpha}_1$ represents the effect of interest.

Construction and analysis of a change score likely represents an attempt to isolate the very same effect by instead estimating the effect of $X_0$ on $\Delta Y = Y_1 - Y_0$ using, for example, a regression model of the form $\widehat{\Delta Y} = \hat{\beta}_0 + \hat{\beta}_1 X_0$. We refer to this analysis as the **change-score analysis**, where $\hat{\beta}_1$ represents the coefficient that is often (mis)interpreted as the effect of interest. Instead of 'standardising' $Y_1$ by $Y_0$, however, the change score treats two separate events (i.e. $Y_0$ and $Y_1$) as one, thereby conflating the causal pathways involved and introducing potential inferential bias.

The degree of discordance between the coefficients of interest in the *follow-up adjusted for baseline analysis* (i.e. $\hat{\alpha}_1$) and the *change-score analysis* (i.e. $\hat{\beta}_1$) will depend on the strength of the association between the baseline exposure $X_0$ and the baseline outcome $Y_0$. Where the association between $X_0$ and $Y_0$ is trivial, the association between $X_0$ and $\Delta Y$ (i.e. $\hat{\beta}_1$) will converge on the association between $X_0$ and $Y_1$ (i.e. $\hat{\alpha}_1$). This would be expected in large, well-conducted randomised experimental studies (i.e. Figure 2A), in which change-score analyses may be used without invoking inferential bias.

*[Insert Figure 2 here]*

However, as the association between $X_0$ and $Y_0$ strengthens – as in non-randomised, non-experimental (i.e. observational) settings – the association between $X_0$ and $\Delta Y$ (i.e. $\hat{\beta}_1$) will be increasingly dominated by the component '$-Y_0$', thereby diverging from the association between $X_0$ and $Y_1$ (i.e. $\hat{\alpha}_1$). Whilst *statistically* unbiased, $\hat{\beta}_1$ may nevertheless invoke inferential bias, since its divergence from $\hat{\alpha}_1$ can be substantial and even sign-reversing. For example, if $X_0$ and $Y_0$ share a strong positive correlation, the negative transformation of $Y_0$ in the change score may dominate a smaller positive correlation between $X_0$ and $Y_1$, resulting in an overall negative association between $X_0$ in $\Delta Y$.

## EXOGENOUS CHANGE VERSUS TOTAL CAUSAL EFFECTS

It may be tempting to conclude that a *follow-up adjusted for baseline analysis* thus represents the best solution for analyses of change in observational data, in which an association between $X_0$ and $Y_0$ is expected. Nevertheless, consideration must also be given to the *direction* of any causal association between $X_0$ and $Y_0$ and the implications for the true quantity of interest.

The randomised experimental setting is unique for ensuring that $X_0$ occurs at the same time or after $Y_0$ by design. This guarantees that all changes in $Y$ that are caused by $X_0$ will be fully realised by the effect of $X_0$ on $Y_1$. In other words, the experimental setting ensures that the effect of $X_0$ on exogenous change in $Y$ is equal to the total causal effect of $X_0$ on $Y_1$. However, this cannot be generalised to all observational settings.

In some non-randomised contexts, such as where the baseline exposure is fast-acting and/or weakly autocorrelated over time, it may be obvious that $X_0$ occurs after $Y_0$, and that the dominant direction of causality therefore flows from $Y_0$ to $X_0$ (Figure 2B). In this setting, the effect of $X_0$ on exogenous change in $Y$ again corresponds to the total causal effect of $X_0$ on $Y_1$, and a *follow-up adjusted for baseline analysis* is appropriate (and necessary) since $Y_0$ is a classical confounder for any effect of $X_0$ on $Y_1$.

However, in many other contexts it is plausible that the baseline exposure causes both the baseline values of the outcome and the follow-up values of the outcome, due to delayed or prolonged causal effects. In such circumstances, the dominant direction of causality flows from $X_0$ to $Y_0$ (Figure 2C), and $X_0$ causes 'changes' in $Y$ due to its effects on *both* $Y_0$ and $Y_1$. In this context, the effect of $X_0$ on exogenous change in $Y$ is arguably less meaningful, since it corresponds only to the direct effect of $X_0$ on $Y_1$. If this effect is sought, then a *follow-up adjusted for baseline analysis* may be appropriate – though such a strategy would involve conditioning on the mediator $Y_0$, which introduces additional methodological challenges.[11, 12] However, if it is the total effect that is sought, then a **follow-up unadjusted for baseline analysis** should be conducted. This would involve constructing, for example, a regression model of the form $\widehat{Y_1} = \hat{\gamma}_0 + \hat{\gamma}_1 X_0$, where $\hat{\gamma}_1$ represents the causal effect of interest.

The choice of whether to adjust for the baseline outcome (i.e. $Y_0$) is therefore *context-dependent*, as it depends upon the hypothesised causal relationship between the baseline exposure and outcome.

## ILLUSTRATIVE EXAMPLE

To illustrate the inferential bias that may be introduced from naïve analyses of change scores, we consider the context of the longitudinal effect of waist circumference ($WC$) on (log-transformed) serum insulin concentration ($IC$) in US adults aged 18-49 years from 2009-2014.[13]

### Methods

Data were simulated to match eight simplified causal scenarios (Figure 3):

1. $IC$ at baseline ($IC_0$) is not directly causally related to $WC$ at baseline ($WC_0$), making it a 'competing exposure' for the effect of $WC_0$ on follow-up $IC$ ($IC_1$).

    A. No unmeasured confounding.

    B. Unmeasured variable ($U$) affecting all three source variables.

2. $IC$ at baseline ($IC_0$) affects $WC$ at baseline ($WC_0$), making it a confounder for the effect of $WC_0$ on follow-up $IC$ ($IC_1$).

    A. No unmeasured confounding.

    B. Unmeasured variable ($U$) affecting all three source variables.

3. $IC$ at baseline ($IC_0$) is affected by $WC$ at baseline ($WC_0$), making it a mediator for the effect of $WC_0$ on follow-up $IC$ ($IC_1$).

    A. No unmeasured confounding.

A+. Unmeasured variable ($U_2$) affecting $IC_0$ and $IC_1$ (i.e. 'mediator-outcome confounding'[12]).

B. Unmeasured variable ($U$) affecting all three source variables.

B+. Unmeasured variable ($U$) affecting all three source variables, and unmeasured variable ($U_2$) affecting $IC_0$ and $IC_1$ (i.e. 'mediator-outcome confounding').

*[Insert Figure 3 here]*

Parameter values were informed by data from the US National Health and Nutrition Examination Survey (NHANES), for the years 2009-2014.[13] The total causal effect of $WC_0$ on $IC_1$ was fixed at 0.200 Log[mmol/L]/dm; when mediated through $IC_0$, this was partitioned into an indirect causal effect of 0.150 Log[mmol/L]/dm and a direct causal effect of 0.050 Log[mmol/L]/dm. Full details of the simulation are provided in Appendix A.

For each scenario, we then conduced three analyses using the resulting data:

1. A *change-score analysis*: $\widehat{\Delta IC} = \hat{\beta}_0 + \widehat{\boldsymbol{\beta}_1} WC_0$.
2. A *follow-up adjusted for baseline analysis*: $\widehat{IC_1} = \hat{\alpha}_0 + \widehat{\boldsymbol{\alpha}_1} WC_0 + \hat{\alpha}_2 IC_0$.
3. A *follow-up <u>un</u>adjusted for baseline analysis*: $\widehat{IC_1} = \hat{\gamma}_0 + \widehat{\boldsymbol{\gamma}_1} WC_0$.

We considered the implications of interpreting the resulting regression coefficient of $WC_0$ (i.e. $\hat{\beta}_1$, $\hat{\alpha}_1$, or $\hat{\gamma}_1$) as the desired causal effect on $IC_1$. Coefficient units (i.e. Log[mmol/L]/dm) are omitted to aid readability.

**Results**

The resulting regression coefficients of $WC_0$ for each of the three methods of analysis for each of the three scenarios are summarised in Table 1.

*[Insert Table 1 here]*

***Scenario 1: Baseline insulin as a 'competing exposure'***

Scenario 1A is analogous to a large, well-conducted randomised experimental study. The association between $WC_0$ and $\Delta IC$ thus consists entirely of the causal effect of $WC_0$ on $IC_1$. Since there is no confounding or mediation by $IC_0$, all methods of analysis provide an unbiased estimate of the total causal effect of $WC_0$ on 'change' in $IC$ ($\hat{\beta}_1 = \hat{\alpha}_1 = \hat{\gamma}_1 = 0.200$).

In Scenario 1B, the association between $WC_0$ and $\Delta IC$ again consists of the causal effect of $WC_0$ on $IC_1$ but this is now confounded by $U$. All three methods of analysis provide a biased estimate of the causal effect of $WC_0$ ($\hat{\beta}_1 = 0.191$, $\hat{\alpha}_1 = 0.203$, $\hat{\gamma}_1 = 0.228$). However, it is worth noting that the *follow-up adjusted for baseline* estimate (i.e. $\hat{\alpha}_1$) is less biased than the *follow-up <u>un</u>adjusted for baseline estimate* (i.e. $\hat{\gamma}_1$), since adjustment for $IC_0$ closes one of the two confounding paths between $WC_0$ and $IC_1$.

### *Scenario 2: Baseline insulin as a confounder*

In Scenario 2A, the association between $WC_0$ and $\Delta IC$ consists of the causal effect of $WC_0$ on $IC_1$ and confounding by $IC_0$. Both the *change-score analysis* and *follow-up <u>un</u>adjusted for baseline analysis* provide biased estimates of the total causal effect of $WC_0$ on $IC_1$ ($\hat{\beta}_1 = 0.119$ and $\hat{\gamma}_1 = 0.351$, respectively). The *follow-up adjusted for baseline analysis* recovers the correct total causal effect ($\hat{\alpha}_1 = 0.200$) because conditioning on $IC_0$ closes the confounding path through $IC_0$.

In Scenario 2B, the association between $WC_0$ and $\Delta IC$ consists of the causal effect of $WC_0$ on $IC_1$ and confounding from both $IC_0$ and $U$. All methods of analysis provide a biased estimate of the total causal effect of $WC_0$ ($\hat{\beta}_1 = 0.114$, $\hat{\alpha}_1 = 0.205$, $\hat{\gamma}_1 = 0.382$), though the *follow-up adjusted for baseline analysis* remains the least biased.

### *Scenario 3: Baseline insulin as a mediator*

In Scenario 3A, the association between $WC_0$ and $\Delta IC$ consists of both the direct causal effect of $WC_0$ on $IC_1$ *and* the indirect causal effect that is mediated through $IC_0$. The *change-score analysis* ($\hat{\beta}_1 = -0.031$) provides a biased estimate of (and is also of the opposite sign to) both the total and direct causal effects of $WC_0$ on $IC_1$. The *follow-up adjusted for baseline analysis* provides an unbiased estimate of the direct causal effect of $WC_0$ on $IC_1$ ($\hat{\alpha}_1 = 0.050$), though the estimate is biased ($\hat{\alpha}_1 = 0.025$) in the presence of mediator-outcome confounding (Scenario 3A+), since conditioning on $IC_0$ opens a confounding path through $U_2$.[12] The *follow-up <u>un</u>adjusted for baseline analysis* provides an unbiased estimate of the total causal effect of $WC_0$ on $IC_1$ ($\hat{\gamma}_1 = 0.200$), which remains robust in the presence of mediator-outcome confounding (Scenario 3A+).

In Scenario 3B, as previously, the association between $WC_0$ and $\Delta IC$ again consists of the direct causal effect of $WC_0$ on $IC_1$ and the indirect causal effect mediated through $IC_0$, but this is now confounded by $U$. The *change-score analysis* remains biased ($\hat{\beta}_1 = -0.031$) and misleading. Both the *follow-up adjusted for baseline analysis* and *follow-up <u>un</u>adjusted for baseline analysis* provide biased estimates of the direct causal effect ($\hat{\alpha}_1 = 0.047$) and total causal effect ($\hat{\gamma}_1 = 0.228$) of $WC_0$, respectively. The bias of the *follow-up adjusted for baseline analysis* is exacerbated ($\hat{\gamma}_1 = 0.015$) in the presence of mediator-outcome confounding (Scenario 3B+) due to conditioning on the collider $IC_0$.

## DISCUSSION

Our study explains why analyses of change scores do not estimate causal effects in observational data. To demonstrate, we explored the ostensibly simple context of analysis of change in an outcome (insulin concentration) with respect to a baseline exposure (waist circumference) for eight different causal scenarios. Misleading coefficients, sometimes of opposite sign to the true total causal effect, were observed in every scenario except where the baseline exposure and baseline outcome were uncorrelated at baseline. Although such independence is plausible, and is indeed actively sought in randomised experimental studies, it is extremely unlikely when the exposure is not assigned randomly. Most analyses of change scores in observational studies are therefore likely to suffer inferential bias, the size of which will vary with the strength and nature of the association between the baseline exposure and baseline outcome.

### Recommendations

Analyses of outcome change scores to estimate causal effects in observational data should be avoided. Instead, the desired causal effect(s) should be formally identified using DAGs and estimated accordingly. We believe this will most often be the total causal effect of the baseline exposure (i.e. $X_0$) on the follow-up outcome (i.e. $Y_1$), as it provides the simplest summary of how changing the exposure would be expected to change future values of the outcome. Where the baseline outcome (i.e. $Y_0$) acts as a mediator for the baseline exposure on the follow-up outcome, the direct effect of $Y_1$ on $Y_1$ may also be sought, but estimating this effect becomes substantially more difficult in the presence of unmeasured mediator-outcome confounding.[11, 12]

**Limitations**

Our simulations were deliberately simplified and made several distributional assumptions that may not be entirely realistic. Multiple variables are likely to confound the true causal effect of waist circumference on insulin concentration. Rather than simulate these individually, we simulated a single summary confounder $U$ for illustrative purposes. The focus of this paper was not however on one specific context; rather, we sought to demonstrate the potential problems with analysing and interpreting change scores in observational studies and the utility of DAGs for exploring and identifying such issues. No inferences should be drawn from our simulations about the assumed causal effect of waist circumference on insulin concentration, which may not exist. We did not consider the additional complications that would result from non-linear relationships, where change scores and linear conditioning for the baseline outcome (e.g. using ANCOVA) would introduce further bias. Where confounding is present and conditioning is required, appropriate parameterisation should be sought to reduce residual confounding.

**Comparison with Lord (1967) and Glymour et al. (2005)**

Scenario 3A, in which the baseline outcome mediates the effect of the exposure on the follow-up outcome, represents the same situation that originally puzzled Lord in 1967.[5] Lord's confusion arose because neither the *change-score analysis* nor the *follow-up adjusted for baseline analysis* seemed to resolve the 'pre-existing' differences in weight at baseline. Using a causal perspective, we can recognise that this 'paradox' occurred for two distinct reasons: (1) the *follow-up adjusted for baseline analysis* does not provide the total causal effect because the baseline outcome is a mediator, and (2) *change-score analyses* do not generally provide meaningful causal effect estimates in observational data. While these points have been individually recognised elsewhere,[7,14] they have yet to be explicitly recognised jointly.

Our conclusion that change scores do not estimate causal effects in non-randomised contexts, including any effect on 'exogenous' change, may explain the divergence between our conclusions and those of Glymour et al. [6] Glymour et al.'s study compares two *change-score analyses*, one *with* and one *without* adjustment for the baseline outcome. In fact, *analyses of change-scores* that adjust for the baseline outcome are equivalent to *follow-up adjusted for baseline analyses*.[15] Thus, the two analytical approaches mirror those involved in Lord's Paradox, and the difference in their results may therefore be at least partly explained by the removal of the '$-Y_0$' component in the adjusted analyses.

## CONCLUSION

Judgements regarding clinical significance and the funding and delivery of treatment are dependent on obtaining meaningful causal effect estimates, and analyses of outcome change scores in non-randomised data do not provide this. Moreover, such analyses may even suggest an 'effect' that is of the opposite sign to the total causal effect. Observational studies that have analysed outcome change scores should therefore be viewed with caution and their recommendations revisited.

Future observational studies should avoid analysing outcome change scores (including 'percentage' change scores, in which the change between baseline and follow-up is expressed as a percentage of the baseline value) and adopt alternative analytical strategies. If the follow-up outcome is not normally-distributed, appropriate transformations and/or non-parametric methods should be preferred to calculating and analysing change scores.[16] Where change scores are preferred for interpretational reasons, they should be calculated from summary data and analysts should be explicit that they were not used in regression or correlation analyses.

15. Laird N. Further Comparative Analyses of Pretest-Posttest Research Designs. *The American Statistician* 1983; **37**: 329-30.

16. Vickers AJ. Parametric versus non-parametric statistics in the analysis of randomized trials with non-normally distributed data. *BMC medical research methodology* 2005; **5**: 35.
16

# TABLES

## Table 1

Regression coefficients (and 95% simulation limits) for the 'effect' of the exposure waist circumference ($WC_0$) on the outcome log insulin concentration (either $\Delta IC$ or $IC_1$) resulting from the three methods of analysis for the eight causal scenarios shown in Figure 3.

| | Regression coefficient for $WC_0$ (Log[mmol/L]/dm) (95% simulation limits) | | | | | | | |
|---|---|---|---|---|---|---|---|---|
| $IC_0$ is: → | Competing exposure | | Confounder | | Mediator | | | |
| Scenario: → | 1A | 1B | 2A | 2B | 3A | 3B | 3A+ | 3B+ |
| **Method of analysis:** ↓ | | | | | | | | |
| Change-score | **0.200** | 0.191 | 0.119 | 0.114 | -0.031 | -0.040 | -0.031 | -0.040 |
| ($\widehat{\Delta IC} = \hat{\beta}_0 + \widehat{\boldsymbol{\beta}_1} WC_0$) | (0.180, 0.221) | (0.172, 0.210) | (0.106, 0.132) | (0.104, 0.123) | (-0.053, -0.009) | (-0.061, -0.019) | (-0.050 -0.012) | (-0.058, -0.023) |
| Follow-up adjusted for baseline | **0.200** | 0.203 | **0.200** | 0.205 | **0.050** | 0.047 | 0.025 | 0.015 |
| ($\widehat{IC_1} = \hat{\alpha}_0 + \widehat{\boldsymbol{\alpha}_1} WC_0 + \hat{\alpha}_2 IC_0$) | (0.182, 0.218) | (0.187, 0.220) | (0.189, 0.211) | (0.199, 0.211) | (0.026, 0.073) | (0.024, 0.071) | (0.005, 0.046) | (-0.005, 0.036) |
| Follow-up <u>un</u>adjusted for baseline | **0.200** | 0.228 | 0.351 | 0.382 | **0.200** | 0.228 | **0.200** | 0.228 |
| ($\widehat{IC_1} = \hat{\gamma}_0 + \widehat{\boldsymbol{\gamma}_1} WC_0$) | (0.174, 0.226) | (0.203, 0.253) | (0.332, 0.369) | (0.366, 0.398) | (0.175, 0.226) | (0.203, 0.253) | (0.174, 0.226) | (0.203, 0.253) |

The total causal effect of $WC_0$ on l$IC_1$ was simulated to be 0.200 Log[mmol/L]/dm; when mediated through $IC_0$, this was partitioned into an indirect causal effect of 0.150 Log[mmol/L]/dm and a direct causal effect of 0.050 Log[mmol/L]/dm. Deviations from these values reflect statistical or inferential bias.

# FIGURES

**Figure 1**

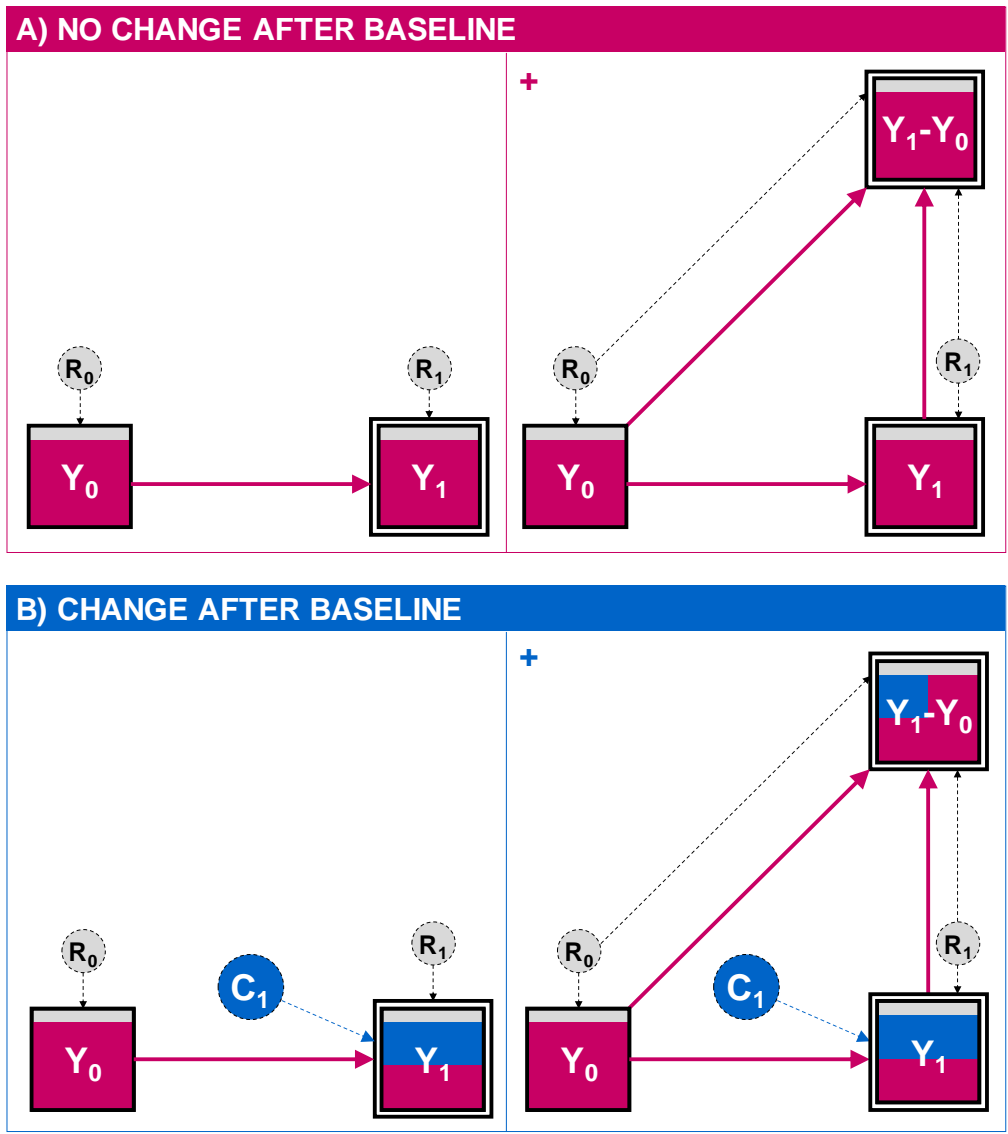

Directed acyclic graphs (DAGs) depicting the relationship between an outcome variable at baseline ($Y_0$) and follow-up ($Y_1$), where the follow-up measure is completely determined. In panel A, the values of $Y_1$ are fully determined by $Y_0$ (and random processes $R_1$), so there exists no exogenous change. In panel B, the values of $Y_1$ are partly determined by $Y_0$ (and random processes $R_1$) and partly determined by exogenous factors representing 'change' ($C_1$). $C_1$, $R_0$, and $R_1$ are depicted as dashed (latent) variables, as they cannot be directly measured. Panels A+ and B+ depict the same causal scenarios as Panels A and B, respectively, but also show the composite change score variable ($Y_1 - Y_0$), which itself is completely determined by $Y_0$ and $Y_1$.

**Figure 2**

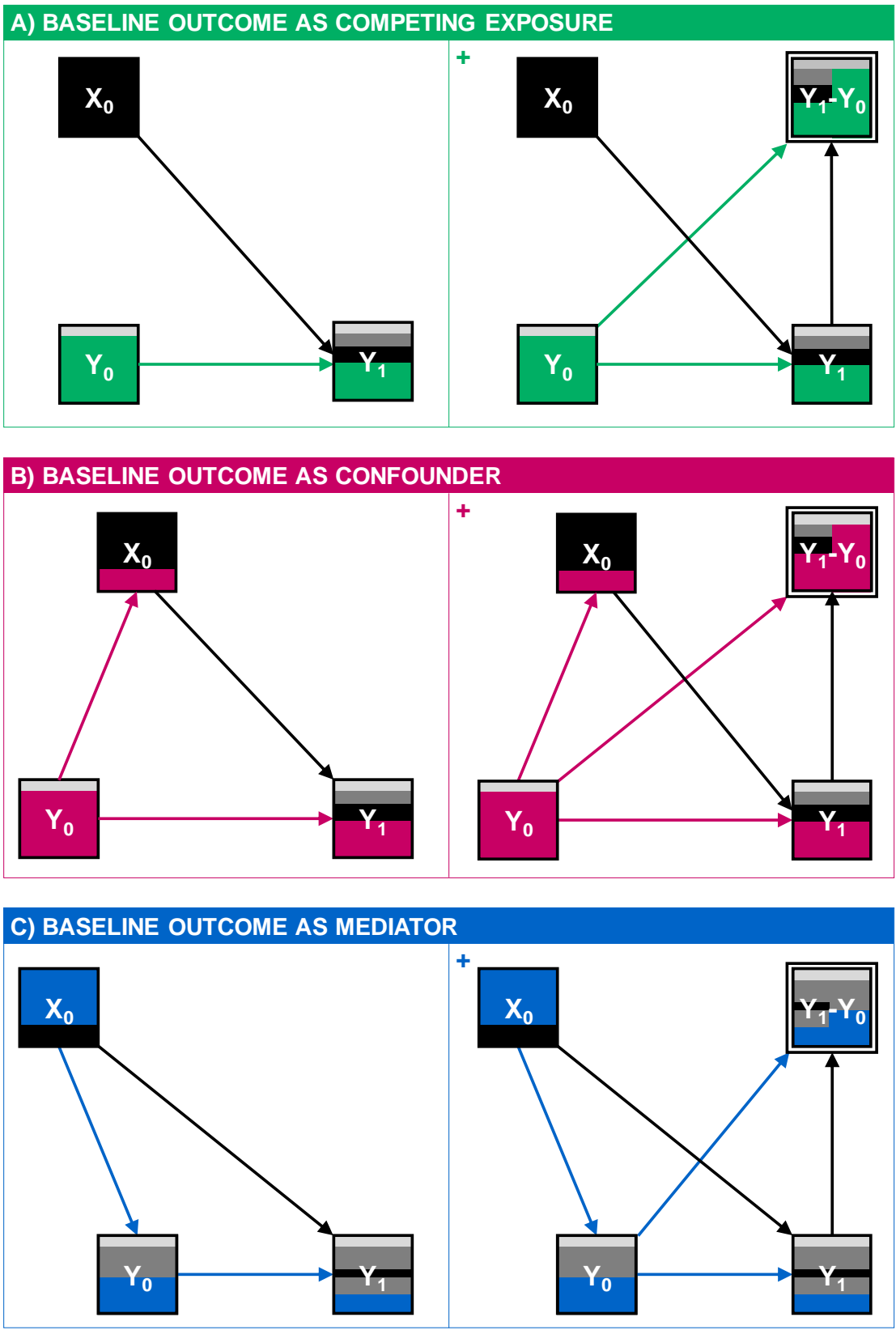

Directed acyclic graphs (DAGs) depicting three causal scenarios for analyses of change in an outcome ($Y$) in relation to a baseline exposure ($X_0$). Panel A represents a well-conducted randomised experimental study, where $X_0$ and $Y_0$ are expected to be unrelated. Panel B represents a scenario where $Y_0$ is a confounder for the effect of $X_0$ on $Y_1$. Panel C represents a scenario where $Y_0$ is a mediator for the effect of $X_0$ on $Y_1$. Panels A+, B+, and C+ depict the same causal scenarios as Panels A, B, and C, respectively, but also depict the composite change score variables ($Y_1 - Y_0$), which are completely determined by $Y_0$ and $Y_1$.

**Figure 3**

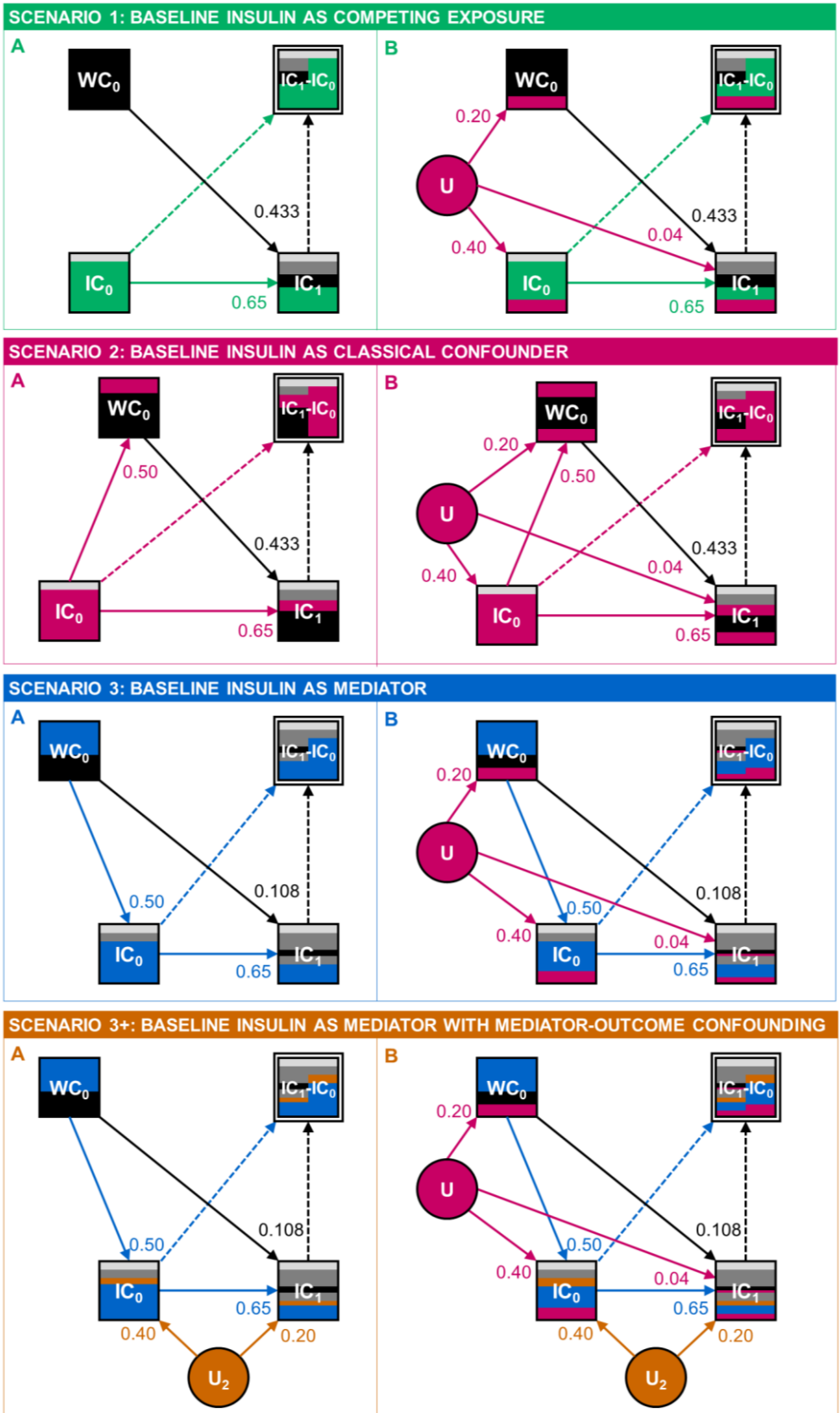

Directed acyclic graphs (DAGs) of the eight simulated scenarios. Waist circumference at baseline ($WC_0$), log insulin concentration at baseline ($IC_0$), log insulin concentration at follow-up ($IC_1$), one or more unobserved confounding variables ($U$), and one or more unobserved mediator-outcome confounding variables ($U_2$) were simulated with the specified path coefficients; for more details see Appendix A. Composite change-score variables ($IC_1 - IC_0$) were derived and are therefore depicted as a double-outlined nodes with dashed incoming arcs, to indicate that these were not simulated. The standardised total causal effect of $WC_0$ on $IC_1$ was fixed at 0.433, as this corresponded with a regression coefficient of 0.200 Log[mmol/L]/dm. When mediated through $IC_0$, the standardised direct effect of $WC_0$ on $IC_1$ was fixed at 0.108, as this corresponded with a regression coefficient of 0.05 Log[mmol/L]/dm.

## ACKNOWLEDGEMENTS

The Authors would like to thank Dr Johannes Textor (Radboud University), Laurie Berrie (University of Leeds), Sarah C Gadd (University of Leeds), and Jake Ellis (University of Leeds) for their helpful comments on previous versions of this manuscript.

## FUNDING

This study received no specific funding. KFA is grateful for funding from the Economic and Social Research Council [grant number ES/J500215/1]. MSG and PWGT are both supported by The Alan Turing Institute [grant number EP/N510129/1].

## AUTHOR CONTRIBUTIONS

MSG conceived the project and with PWGT and KFA designed the study. PWGT, KFA, GTHE, and MSG were involved in designing, testing, conducting, and/or interpreting the simulations. PWGT conducted the final data analysis and with KFA drafted the report. All authors critically reviewed the manuscript. All authors read and approved the final manuscript before submission. PWGT accepts full responsibility for the work and conduct of the study, had full access to the data, and controlled the decision to publish.

## CONFLICT OF INTEREST

The Authors declare that there are no conflicts of interest.

# APPENDIX A

**Data simulation**

Data were simulated to match the eight causal scenarios depicted in Figure 3, with parameter values and path coefficients informed by data from the US National Health and Nutrition Examination Survey (NHANES), for the years 2009-2014 (Table A.1).[17]

Since insulin concentration ($IC$) appears log-normally distributed,[18] we simulated and analysed $IC$ in log-transformed form. For each scenario, multivariate normal data with values for baseline $IC$ ($IC_0$), follow-up $IC$ ($IC_1$), and baseline $WC$ ($WC_0$) were simulated in a sample of 1000 participants using 'dagitty' v.0.2.2 in R 3.4.0.[19,20] The simulated mean (SD) for $IC_0$ was 4.00 Log[mmol/L] (0.74) and 4.20 Log[mmol/L] (0.74) for $IC_1$, representing a notional 5% increase from baseline to follow-up. The simulated mean (SD) for $WC_0$ was 9.5dm (1.6). Path coefficients between these variables were selected to reflect the observed cross-sectional correlation of $r \approx 0.6$ (Table A.I).[17] The path coefficient between $IC_0$ and $IC_1$ was simulated as 0.65, to represent strong but imperfect determination over time. In some scenarios, one or more unobserved confounding factors were represented by $U$, which was simulated to introduce a confounded correlation of $r \approx 0.08$ between $WC_0$ and both $IC_0$ and $IC_1$. The total causal effect of $WC_0$ on $IC_1$ was fixed at 0.2 Log[mmol/L]/dm, equivalent to a direct path coefficient of 0.433. When mediated through $IC_0$, this was partitioned into an indirect causal effect of 0.15 Log[mmol/L]/dm and a direct causal effect of 0.05 Log[mmol/L]/dm. For illustrative purposes, we also simulated mediator-outcome confounding ($U_2$) averaging $r \approx 0.08$ between $IC_0$ and $IC_1$. For each scenario, simulations were repeated 10,000 times[21] and the regression coefficients for $WC_0$ stored. Median values are reported with their 95% simulation limits (2.5 and 97.5 centile estimates from the 10,000 samples).

**Table A.1**

|  | Mean (SD) | | | |
|---|---|---|---|---|
|  | As measured by NHANES | | | Simulated |
|  | **2009-2010** | **2011-2012** | **2013-2014** |  |
| **Waist circumference (dm)** | 9.50 (1.58) | 9.42 (1.61) | 9.52 (1.65) | 9.50 (1.60) |
| **Insulin concentration (Log[mmol/L])** | 4.20 (0.70) | 4.08 (0.74) | 3.98 (0.77) | Baseline: 4.00 (0.74) |
|  |  |  |  | Follow-up: 4.20 (0.74) |
| **Pearson correlation (ρ)** | 0.58[a] | 0.58[a] | 0.60[a] | 0.50 - 0.60 |

[a] Between waist circumference (WC) and log insulin concentration (IC).